\pdfoutput=1

\documentclass[twocolumn,aps,prl,superscriptaddress,showpacs]{revtex4}

\usepackage{graphicx,float}
\usepackage{epsfig}
\usepackage{epstopdf}

\usepackage{amsmath}
\usepackage{amssymb}
\usepackage{bm}
\usepackage{bbm}
\usepackage{dsfont}
\usepackage{bbold}
\usepackage{multirow}
\usepackage{color}
\usepackage[usenames,dvipsnames]{xcolor}
\usepackage[colorlinks=true,citecolor=blue,urlcolor=blue]{hyperref}

\usepackage{hyperref}
\hypersetup{colorlinks=true,linkcolor=blue,citecolor=blue,urlcolor=blue}

\newcommand{\be}{\begin{equation}}
\newcommand{\ee}{\end{equation}}

\newcommand{\sket}[1]{{\ensuremath{\lvert#1\rangle}}}
\newcommand{\lket}[1]{{\ensuremath{\left\lvert#1\right\rangle}}}
\newcommand{\ket}[1]{\if@display\lket{#1}\else\sket{#1}\fi}

\newcommand{\sbra}[1]{{\ensuremath{\langle#1\rvert}}}
\newcommand{\lbra}[1]{{\ensuremath{\left\langle#1\right\rvert}}}
\newcommand{\bra}[1]{\if@display\lbra{#1}\else\sbra{#1}\fi}

\newcommand{\sbraket}[2]{{\ensuremath{\langle#1\rvert#2\rangle}}}
\newcommand{\lbraket}[2]{{\ensuremath{\left\langle#1\!\left\rvert\vphantom{#1}#2\right.\!\right\rangle}}}
\newcommand{\braket}[2]{\if@display\lbraket{#1}{#2}\else\sbraket{#1}{#2}\fi}

\newcommand{\sketbra}[2]{{\ensuremath{\lvert #1\rangle\!\langle #2\rvert}}}
\newcommand{\lketbra}[2]{{\ensuremath{\left\lvert #1\right\rangle\!\!\left\langle #2\right\rvert}}}
\newcommand{\ketbra}[2]{\if@display\lketbra{#1}{#2}\else\sketbra{#1}{#2}\fi}

\begin{document}


\title{Postselection-Loophole-Free Bell Test Over an Installed Optical Fiber Network}


\date{\today}

\author{Gonzalo~Carvacho}
\affiliation{Departamento de F\'isica, Universidad de Concepci\'on, 160-C Concepci\'on, Chile}
\affiliation{Center for Optics and Photonics, Universidad de Concepci\'on, Casilla 4016, Concepci\'on, Chile}
\affiliation{MSI-Nucleus for Advanced Optics, Universidad de Concepci\'on, Concepci\'on, Chile}

\author{Jaime~Cari\~ne}
\affiliation{Center for Optics and Photonics, Universidad de Concepci\'on, Casilla 4016, Concepci\'on, Chile}
\affiliation{MSI-Nucleus for Advanced Optics, Universidad de Concepci\'on, Concepci\'on, Chile}
\affiliation{Departamento de Ingenier\'ia El\'ectrica, Universidad de Concepci\'on,160-C Concepci\'on, Chile}

\author{Gabriel~Saavedra}
\affiliation{Center for Optics and Photonics, Universidad de Concepci\'on, Casilla 4016, Concepci\'on, Chile}
\affiliation{MSI-Nucleus for Advanced Optics, Universidad de Concepci\'on, Concepci\'on, Chile}
\affiliation{Departamento de Ingenier\'ia El\'ectrica, Universidad de Concepci\'on,160-C Concepci\'on, Chile}

\author{\'Alvaro~Cuevas}
\affiliation{Departamento de F\'isica, Universidad de Concepci\'on, 160-C Concepci\'on, Chile}
\affiliation{Center for Optics and Photonics, Universidad de Concepci\'on, Casilla 4016, Concepci\'on, Chile}
\affiliation{MSI-Nucleus for Advanced Optics, Universidad de Concepci\'on, Concepci\'on, Chile}

\author{Jorge~Fuenzalida}
\affiliation{Departamento de F\'isica, Universidad de Concepci\'on, 160-C Concepci\'on, Chile}
\affiliation{Center for Optics and Photonics, Universidad de Concepci\'on, Casilla 4016, Concepci\'on, Chile}
\affiliation{MSI-Nucleus for Advanced Optics, Universidad de Concepci\'on, Concepci\'on, Chile}

\author{Felipe~Toledo}
\affiliation{Departamento de F\'isica, Universidad de Concepci\'on, 160-C Concepci\'on, Chile}
\affiliation{Center for Optics and Photonics, Universidad de Concepci\'on, Casilla 4016, Concepci\'on, Chile}
\affiliation{MSI-Nucleus for Advanced Optics, Universidad de Concepci\'on, Concepci\'on, Chile}

\author{Miguel~Figueroa}
\affiliation{Center for Optics and Photonics, Universidad de Concepci\'on, Casilla 4016, Concepci\'on, Chile}
\affiliation{Departamento de Ingenier\'ia El\'ectrica, Universidad de Concepci\'on,160-C Concepci\'on, Chile}

\author{Ad\'an~Cabello}
\affiliation{Departamento de F\'isica Aplicada II, Universidad de Sevilla, E-41012, Sevilla, Spain.}

\author{Jan-\AA ke~Larsson}
\affiliation{Institutionen f\"{o}r Systemteknik, Link\"{o}pings Universitet, 581 83 Link\"{o}ping, Sweden}

\author{Paolo~Mataloni}
\affiliation{Dipartimento de Fisica, Sapienza Universit\`a di Roma, Piazzale Aldo Moro 5, Roma I-00185, Italy.}

\author{Gustavo~Lima}
\affiliation{Departamento de F\'isica, Universidad de Concepci\'on, 160-C Concepci\'on, Chile}
\affiliation{Center for Optics and Photonics, Universidad de Concepci\'on, Casilla 4016, Concepci\'on, Chile}
\affiliation{MSI-Nucleus for Advanced Optics, Universidad de Concepci\'on, Concepci\'on, Chile}

\author{Guilherme B.~Xavier}
\email{gxavier@udec.cl}
\affiliation{Center for Optics and Photonics, Universidad de Concepci\'on, Casilla 4016, Concepci\'on, Chile}
\affiliation{MSI-Nucleus for Advanced Optics, Universidad de Concepci\'on, Concepci\'on, Chile}
\affiliation{Departamento de Ingenier\'ia El\'ectrica, Universidad de Concepci\'on,160-C Concepci\'on, Chile}

\email{gxavier@udec.cl}

\begin{abstract}
Device-independent (DI) quantum communication will require a loophole-free violation of Bell inequalities. In typical scenarios where line-of-sight between the communicating parties is not available, it is convenient to use energy-time entangled photons due to intrinsic robustness while propagating over optical fibers. Here we show an energy-time Clauser-Horne-Shimony-Holt Bell inequality violation with two parties separated by 3.7 km over the deployed optical fiber network belonging to the University of Concepci\'on in Chile. Remarkably, this is the first Bell violation with spatially separated parties that is free of the post-selection loophole, which affected all previous in-field long-distance energy-time experiments. Our work takes a further step towards a fiber-based loophole-free Bell test, which is highly desired for secure quantum communication due to the widespread existing telecommunication infrastructure.

\end{abstract}


\pacs{03.67.Dd, 03.67.Hk}


\maketitle


{\em Introduction.---} Observing a loophole-free Bell inequality violation on composite systems, such as pairs of entangled photons \cite{Bell}, is crucial for secure quantum communication \cite{Ekert91, GisinRMP}. In order to take advantage of the existing optical communication links \cite{Frohlich13}, transmission through optical fibers is often desirable. For this task, energy-time entanglement has been a widely employed choice for many years, due to its resilience against decoherence linked to propagation in optical fibers \cite{Rarity94, Tittel98, Brendel99, Tittel00, Marcikic04, Salart08, Dynes09, Inagaki13}. One promising approach for the construction of unbreakable cryptographic systems is called ``device-independent'' quantum key distribution (QKD), whose security is certified even if the devices, or the quantum systems being measured, are not characterized \cite{Acin07}. This is true provided that there is a loophole-free Bell inequality violation. However, if loopholes are present in the experimental implementation, security can no longer be guaranteed \cite{Larsson14}. Although considerable progress has taken place in the last few years, and Bell inequality violations have been observed free of some loopholes \cite{Weihs98, Rowe01, Monroe08, Zeilinger13, Kwiat13, Cuevas13, Smith2012, Bennet2012, Wittmann2012}, the goal of a violation free of all loopholes is still missing.

\begin{figure}[t]
\centering
\includegraphics[width=0.5 \textwidth]{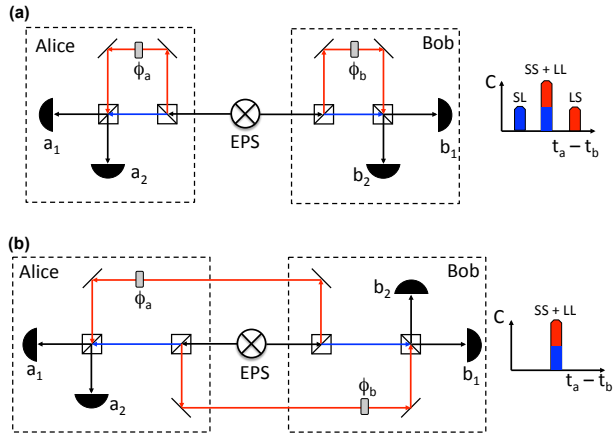}
\caption{(color online). Energy-time Bell setups. (a) Franson's scheme: the entangled photon source (EPS) sends a photon to each party (Alice and Bob), each possessing an UMZI interferometer composed of a short path (S - blue) and a long one (L - red). Each path is superposed on a beamsplitter, whose two output ports are connected to single-photon detectors $a_i$ and $b_i$ with $i\in \{0,1\}$. SL and LS events are discarded from the detection results, as they are both distinguishable due to different detection times [inset of Fig.~\ref{fig1}(a)], where $t_x$ corresponding to the detection time of a single-photon, and $x \in \{a,b\}$. C corresponds to the rate of coincidence detection. (b) Hug configuration. This scheme has a different geometrical structure, in which the LS and SL combinations are always routed to the same party, and therefore no classical communication between Alice and Bob is needed to discard events. Coincidence events across Alice and Bob's detectors can only occur for SS and LL detections, thereby removing the post-selection loophole [inset of Fig.~\ref{fig1}(b)].}\label{fig1}
\end{figure}

The standard experimental configuration to observe a Bell inequality violation with energy-time entangled photons, adopted by most implementations until today, was proposed in 1989 by Franson \cite{Franson89}. It consists of a source creating photon pairs at unknown times, with each photon from each pair transmitted over communication channels to two unbalanced Mach-Zehnder interferometers (UMZIs) as shown in Fig.~\ref{fig1}(a). In each UMZI, the corresponding single-photon can randomly take one of two possible paths, long (L) or short (S), which are then superposed on a beamsplitter. The L-S path difference in each UMZI is the same within the coherence length of the pair, and therefore four possibilities can be detected in coincidence: LL, SS, LS, and SL. The coincidence window is adjusted (depending on the length of the L-S path difference) such that the LS and SL events are discarded. After this step, the remaining coincidences are indistinguishable in principle and a high visibility two-photon interference pattern emerges. In this condition the following entangled state is generated: $\ket{\Phi}=1/\sqrt{2}(\ket{SS}+e^{(\phi_a-\phi_b)}\ket{LL})$, where $\phi_a$ and $\phi_b$ are phase shifts within the long arm of Alice and Bob's respective UMZIs.

Unfortunately, Franson's configuration has an inherent loophole as the choice of local settings may influence the events that are selected and discarded. This is known as the post-selection loophole, and is discussed at length in a recent review paper \cite{Jogenfors14}. Recovering a test of local realism in
this configuration requires extra assumptions on the physical system used \cite{Franson1, Franson2}, essentially requiring verification of particle-like properties of photons inside the interferometers, something that cannot be done. A local hidden variable model for this configuration was presented in Ref.~\cite{Aerts99}, and the possibility of a Trojan-horse attack exploiting this loophole was described in Ref.~\cite{Larsson02}. To emphasize the grave consequences of this flaw, a blinding attack was recently demonstrated on the Franson scheme, effective even with near-perfect efficiency detectors \cite{Jogenfors14b}.

In this work, we carry out an energy-time Bell test free of the post-selection loophole between two parties that are spatially separated over 1.5 km, with a total propagation distance of 3.7 km through the deployed optical fiber network of the University of Concepci\'on in Chile. A solution to completely remove the post-selection loophole is to modify the geometry of the setup to a cross-linked interferometric scheme, known as the hug configuration \cite{CRVMM09} [see Fig.~\ref{fig1}(b)]. Although it is experimentally more challenging, it has been successfully demonstrated as a proof-of-principle \cite{LVCCM10} and then up to distances of 1 km of spooled optical fibers \cite{Cuevas13}. A major technological challenge is successfully overcome here, compared to our previous demonstrations \cite{LVCCM10, Cuevas13}, by actively stabilizing a 3.7 km fiber-optical UMZI across the campus of the university. Our new results (i) extend those initial efforts in cementing the hug configuration as practical and (ii) demonstrates that energy-time entanglement can actually be used to implement truly secure communications over the existing telecommunication infrastructure. This work opens up the path towards device-independent QKD across deployed optical fiber networks.

\begin{figure}[t]
\centering
\includegraphics[width=0.5 \textwidth]{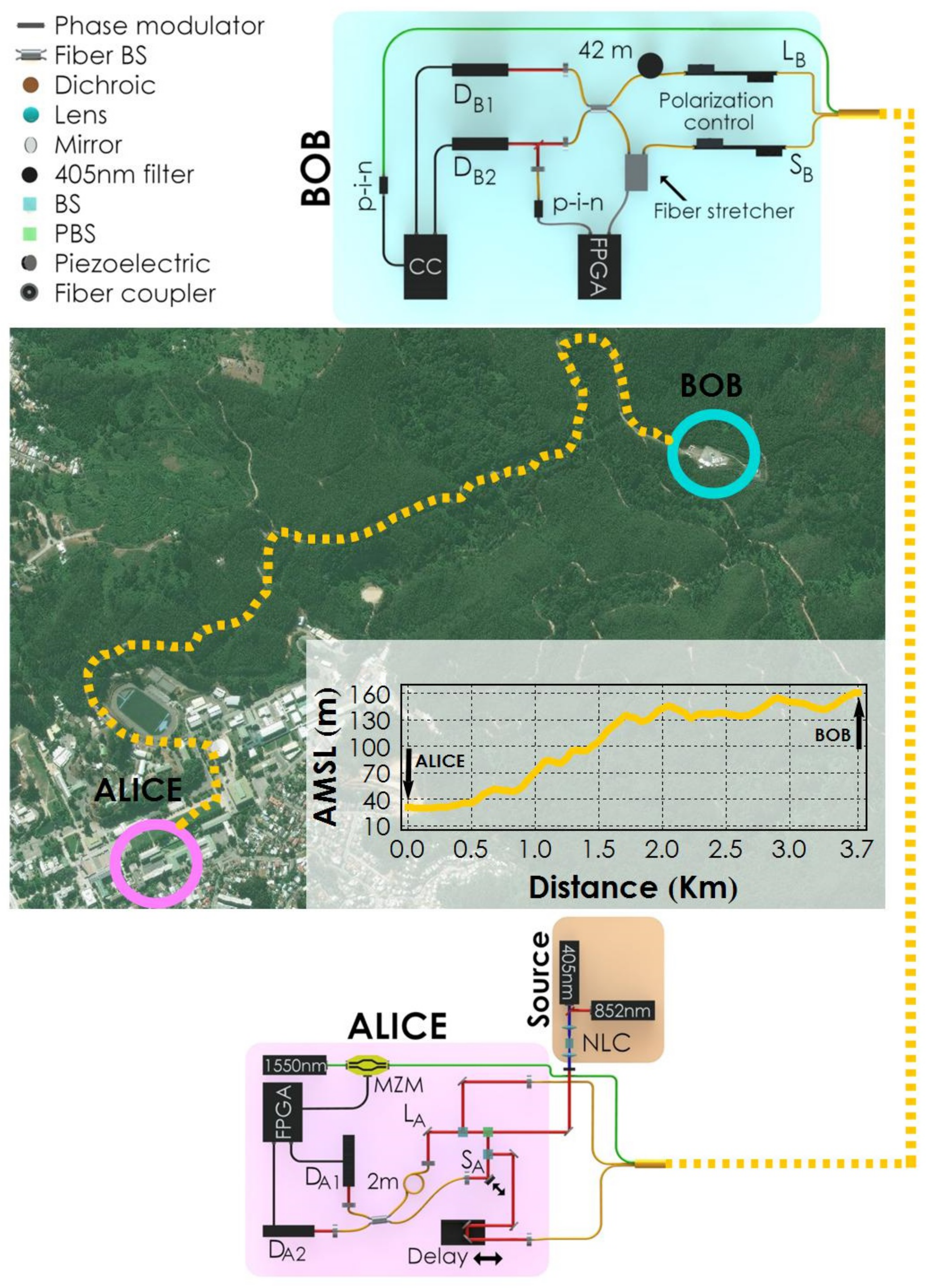}
\caption{Experimental setup. Alice and Bob are placed at remote locations throughout the campus of the University of Concepci\'on in Chile. Alice and the source are located at the Faculty of Engineering, while Bob is located at the Transportable Integrated Geodesic Observatory (TIGO). They are connected through 3.7 km of optical fibers belonging to the network of the university, represented as a dashed yellow line in the overview map and the schematics of the experiment. The map inset shows the above mean sea level (AMSL) height variation between Alice and Bob, clearly showing no direct line-of-sight is available. The spatial separation between Alice and Bob is approximately 1.5 km. Four dark fibers are used in the demonstration: two for the short and long arms connecting the source to Bob ($S_B$ and $L_B$), the third is used to provide timing information for the coincidence electronics placed within Bob (see text for details), and the fourth is used for remote operation of Bob's station (not shown for clarity). Imagery \copyright 2015 DigitalGlobe, Map data \copyright 2015 Google.
\label{fig2}}
\end{figure}

{\em Experiment.---}In our experiment, the source is located within Alice's laboratory where pairs of degenerate 806 nm type-II entangled photons are created in a periodically poled potassium titanyl phosphate (ppKTP) 20 mm long crystal, when pumped with 4 mW of optical power from a long coherence (greater than 20 m) 403 nm continuous wave pump diode laser (Fig.\ref{fig2}). They are deterministically split with a polarizing beamsplitter, and each photon enters the cross-linked interferometers of the hug configuration. The two arms connecting the source to Alice ($S_A$ and $L_A$) are coupled to single-mode optical fibers and are connected to a single-mode fiber bidirectional coupler (single-mode for 780 nm and above), for optimal spatial superposition. Each output of the coupler is then connected to a single-photon avalanche detector in free-running mode, with an overall detection efficiency of $\sim 60\%$.  In the short path, a mirror mounted on a piezo-electric element (P) was used to implement the local phase shift $\phi_a$. The difference between the short and long arms is of 2 m, while the length of the short arm is 1 m in length.


The experiment consists of three main challenges. The first one is the implementation of the long UMZI throughout the campus of the university connecting the source to Bob. We take advantage of the fact that installed optical cables are typically comprised of many individual fibers, and use two of them for the UMZI.  In our case, the arms connecting the source to Bob ($S_B$ and $L_B$), are therefore two 3.7 km standard telecommunication dark optical fibers (SMF-28) inside a cable deployed throughout the campus [see Fig.(\ref{fig2})]. These fibers actually support two linearly polarized modes of propagation \cite{Saleh} at 806 nm, but fortunately, the higher order mode can be easily filtered out by Bob with single-mode fiber-optical components \cite{Meyer-Scott10}. The 3.7 km distance was determined with an optical time-domain reflectometer (OTDR). The difference between the propagation vs. straight line distance can be expected in real installations, due to the locations of the different access and routing points.

The link loss in the UMZI arms is 17 dB, which is higher than the expected 12 dB for a length close to 4 km at 806 nm. This is due to higher-than-usual connector/splicing losses in the link as seen from the OTDR measurement. Optical filters placed in front of the detectors (1 nm passband) with 90\% transmission were used to partially compensate these losses. We have also designed coincidence electronics based on a field programmable gate array (FPGA) platform, in order to implement 1 ns coincidence windows, reducing accidental counts, compared to \cite{Cuevas13} where 4 ns windows were employed. We obtained on the order of 300.000 counts/s at either of Alice's detectors, 9000 counts/s at Bob's and a total of 40 mean coincidence counts/s.

The second experimental challenge is the random phase drift in the long UMZI in the field. The hug configuration requires at least one UMZI that is as long as the propagation distance between the communicating parties, and thus being subjected to random environmentally-induced phase drifts \cite{Minar08}. In our experiment, 1.5 km of the 3.7 km source-to-Bob optical cable is hanging through poles, therefore allowing us to demonstrate the Bell test experiment covering real-life situations. To deal with such phase fluctuations, we extended a technique previously demonstrated in \cite{Xavier11, Cuevas13}. At Bob's station, a piezo-electric fiber stretcher (FS) composed of 40 m single-mode optical fiber is placed in the $S_B$ arm, which is used by an electronic control system to actively compensate the environmental phase fluctuations in the interferometer. The electronics are based on another FPGA circuit, running a proportional-integral-derivative (PID) control algorithm, and the total bandwidth of the control system is around 5 kHz. The feedback to the control is provided by an additional reference long-coherence ($> 20$ m) laser at 852 nm, which propagates throughout the entire setup. The optical control signal is detected at Bob by a p-i-n photodiode after being split with a dichroic mirror following one of the output ports of the interferometer (Fig.~\ref{fig2}).

Besides the environmentally-induced phase drift issue, an additional difficulty is that on a long-distance installed aerial cable, strong power fluctuations at 852 nm are observed with SMF-28-single-mode fiber due to the dynamic speckle pattern as a result of the multimode propagation \cite{Saleh}. These fluctuations jeopardized  the control system since this dynamic pattern caused slowly varying power fluctuations, which are observed at the p-i-n photodiode's output feeding the control system. Moreover, these fluctuations are different for each arm of our UMZI, connecting the source to Bob. As an effect one has a time-varying visibility on the interference pattern of the 852 nm control signal. This time-dependent fluctuation in the visibility is due to the difference in each arm's optical intensity. Without any further action phase-locking the interferometer at any desired intensity on the interference pattern is not possible, since the control set point, depending on the visibility change, may find itself outside of the pattern, causing the control system to oscillate randomly. The adopted solution to this problem is to remove any d-c (offset) component from the electrical output generated from the p-i-n photodiode and lock the interferometer in quadrature (a relative phase of 45$^{\circ}$ or -45$^{\circ}$). 

Finally it is necessary to guarantee the indistinguishability condition in the experiment ($L_A - S_A = L_B - S_B$ within the coherence length of the single-photons, which is of the order of 1 mm) and synchronise Alice and Bob's detectors for coincidence detection. A 15 cm optical delay line is used at the $L_B$ arm to adjust the setup, such that the indistinguishability condition is achieved. This point is reached once two-photon interference fringes are detected between Alice and Bob's detectors, and in our case it happened when $L_A - S_A = L_B - S_B$ were balanced to within 1 mm of path difference. The synchronisation between Alice and Bob's detectors, for coincidence measurements, is provided by sending optical pulses driven by detection within Alice to Bob through a third optical fiber (within the employed optical cable). Each detected photon at either one of Alice's detectors drives a fiber pigtailed Mach-Zehnder optical intensity modulator (MZM), generating optical pulses from a semiconductor distributed feedback (DFB) 1550 nm laser. These pulses are detected by Bob through a dedicated p-i-n photodiode, and fed to the coincidence electronics. Bob's interferometer is phase-locked, and then by performing coincidence measurements, the delay line is properly adjusted in the indistinguishability position \cite{Cuevas13}.

Measurements were then taken in coincidence across all detector combinations with Bob's relative phase $\phi_b$ fixed. All results were taken directly from the raw data (i.e., without accidental subtraction). Alice's piezo-mounted mirror (P) is slowly modulated implementing the phase shift $\phi_a$, generating the results from Fig. 3. From this data we can calculate the value $S$ of the violation of the Clauser-Horne-Shimony-Holt (CHSH) inequality [37], with the extra assumption of symmetry in the experiment [38], since Bob's relative phase had to be fixed during the experiment. In this case, we proceed to calculate the CHSH violation as $S = 3E(\phi_a,\phi_b) - E(\phi^{\prime}_a,\phi_b)$. $E(\phi_a, \phi_b) = P_{11}(\phi_a, \phi_b) + P_{22}(\phi_a, \phi_b) - P_{12}(\phi_a, \phi_b) - P_{21}(\phi_a, \phi_b)$, with $P_{xy}$, corresponding to the probability of obtaining a coincidence detection at Alice's $x$ and Bob's $y$ detectors for a specific setting $\phi_a, \phi_b$. For the maximum violation, $\phi_a = \pi/4$, $\phi_a' = -\pi/4$, with $\phi_b$ fixed at $\pi/4$.  From the raw data from Fig. 3 we calculate the Bell value $S = 2.32 \pm 0.11$, violating the CHSH inequality by over 2.94 standard deviations, with an average visibility of $82.10 \pm 3.87\%$. The deviations from the ideal cosine-squared interference pattern comes mainly from the fact that Alice's interferometer is not actively stabilized, generating small drifts to the pattern.


\begin{figure}[t]
\centering
\includegraphics[width=0.45 \textwidth]{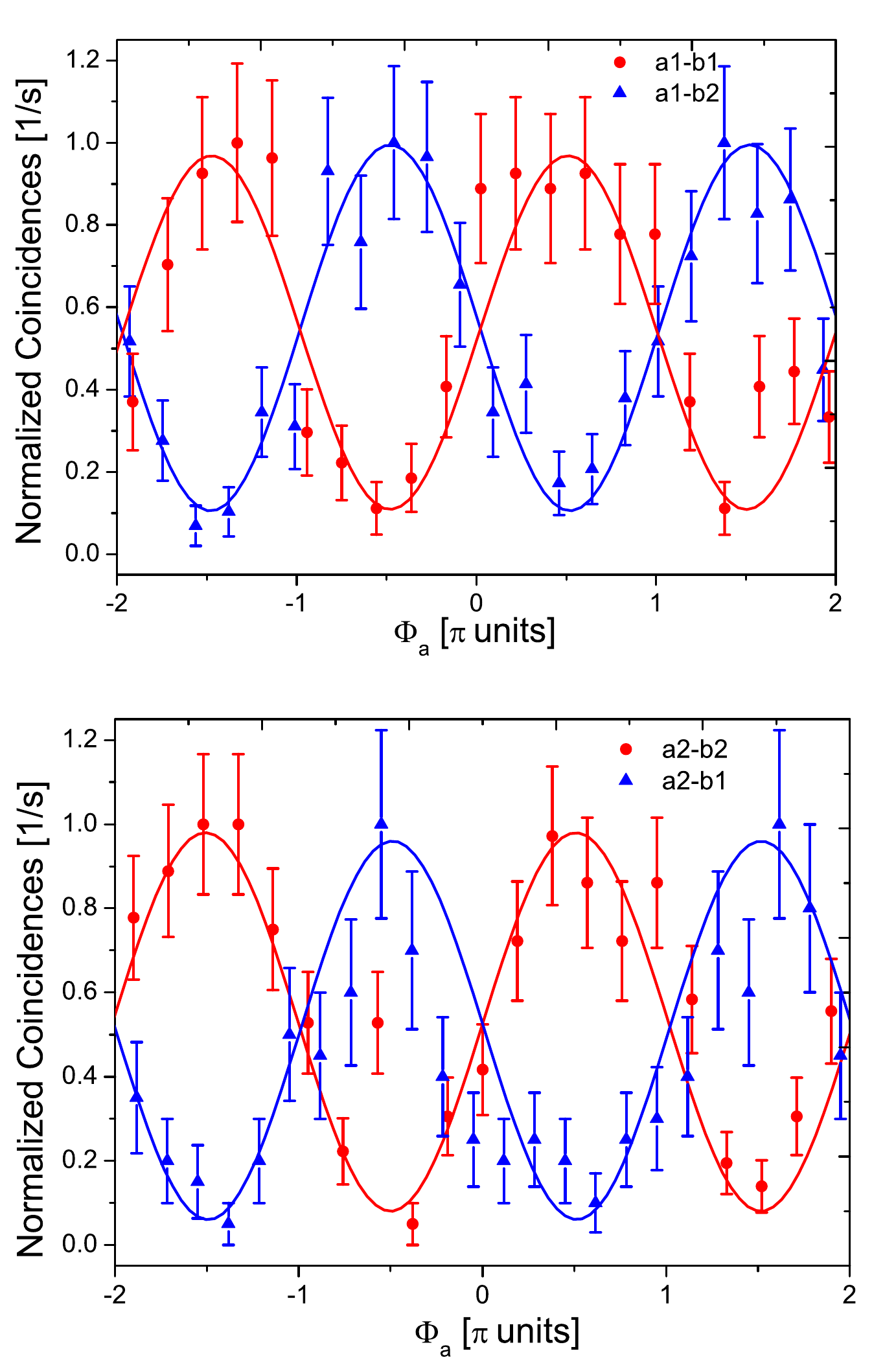}
\caption{Experimental results. Normalized coincidence measurements are shown across Alice's (above) and Bob's (below) detectors. The interference curves are obtained through a variation of the relative phase $\phi_a$ in Alice's interferometer through the piezo-mounted mirror (P). All measurements displayed are without accidental detection subtraction. \label{fig3}}
\end{figure}


{\it Conclusions.---}The DI approach to quantum communications is of major practical relevance because only a few assumptions on the functioning of the employed devices are required \cite{Acin07}. Unconditional security is possible provided there is an observation of a loophole-free Bell inequality violation. During the last years there have been important advances towards this accomplishment \cite{Weihs98, Rowe01, Monroe08, Zeilinger13, Kwiat13, Cuevas13, Smith2012, Bennet2012, Wittmann2012}. In this work we present another step towards fiber-based loophole-free Bell tests, which are specially relevant for DI quantum communication due to the existing optical telecommunication networks. The standard experimental configuration to observe a Bell inequality violation with energy-time entangled photons, adopted by all previous in-field implementations, was proposed in 1989. Nowadays, however, it is well recognised that it has an intrinsic loophole, called post-selection loophole \cite{Aerts99, CRVMM09, Jogenfors14}, that can be exploited for Trojan-horse attacks even with perfect detection efficiencies \cite{Larsson02}. The viability of such attacks has been recently demonstrated experimentally \cite{Jogenfors14b}. Here, we adopt a new geometrical configuration, known as the hug configuration introduced in \cite{CRVMM09}, to show a post-selection loophole-free Bell inequality violation over a total distance of 3.7 kms spread across the campus of the University of Concepci\'on.

In spite of the fact of the more stringent experimental requirements of the hug configuration, our results consolidate it as practical and demonstrates that energy-time entanglement is actually useful for the realization of truly secure communications over the existing telecommunication infrastructure. Future improvements over the electronic systems, operation at telecom wavelengths and the use of more efficient detectors, will allow new possibilities for simultaneously closing other loopholes. As technology matures, fiber-based loophole-free Bell tests will become a new platform for secure quantum communication over the widespread fiber-optical telecommunication infrastructure.


{\it Acknowledgements.---}We would like to thank the LIDAR-CEFOP group, located at TIGO, for assistance in the experiment. This work was supported by the grants CONICYT PFB08-024, Milenio RC130001, FONDECYT (grants No.\ 11110115, 1150101, 1120067, and 1151278), Project No.\ FIS2011-29400 (MINECO, Spain) with FEDER funds, the FQXi large grant project ``The Nature of Information in Sequential Quantum Measurements,'' the program Science without Borders (CAPES and
CNPq, Brazil), project FP7-ICT-2011-9- 600838 (QWAD--Quantum Waveguides Application and Development) and CENIIT at Link\"{o}ping University. G. C., J.C., G.S. and A. Cuevas acknowledge the financial support of CONICYT.




\begin{thebibliography}{000}

\bibitem{Bell}
 J. S. Bell,
 {\em Speakable and Unspeakable in Quantum Mechanics: Collected Papers on Quantum Philosophy} (Cambridge University Press, Cambridge, 2004).

\bibitem{Ekert91}
 A. K. Ekert,
 \href{http://dx.doi.org/10.1103/PhysRevLett.67.661}{Phys. Rev. Lett. \textbf{67}, 661 (1991).}

\bibitem{GisinRMP}
 N. Gisin, G. Ribordy, W. Tittel, and H. Zbinden,
 \href{http://dx.doi.org/10.1103/RevModPhys.74.145}{Rev. Mod. Phys. \textbf{74}, 145 (2002).}

\bibitem{Frohlich13}
 B. Fr\"{o}hlich, J. F. Dynes, M. Lucamarini, A. W. Sharpe, Z. Yuan, and A. J. Shields,
 \href{http://www.nature.com/nature/journal/v501/n7465/full/nature12493.html}{Nature (London) \textbf{501}, 69 (2013).}

\bibitem{Rarity94}
 P. R. Tapster, J. G. Rarity, and P. C. M. Owens,
 \href{http://dx.doi.org/10.1103/PhysRevLett.73.1923}{Phys. Rev. Lett. \textbf{73}, 1923 (1994).}

\bibitem{Tittel98}
 W. Tittel, J. Brendel, H. Zbinden, and N. Gisin,
 \href{http://dx.doi.org/10.1103/PhysRevLett.81.3563}{Phys. Rev. Lett. \textbf{81}, 3563 (1998).}

\bibitem{Brendel99}
 J. Brendel, N. Gisin, W. Tittel, and H. Zbinden,
 \href{http://dx.doi.org/10.1103/PhysRevLett.82.2594}{Phys. Rev. Lett. \textbf{82}, 2594 (1999).}

\bibitem{Tittel00}
 W. Tittel, J. Brendel, H. Zbinden, and N. Gisin,
 \href{http://dx.doi.org/10.1103/PhysRevLett.84.4737}{Phys. Rev. Lett. \textbf{84}, 4737 (2000).}

\bibitem{Marcikic04}
 I. Marcikic, H. de Riedmatten, W. Tittel, H. Zbinden, M. Legr\'e, and N. Gisin,
 \href{http://dx.doi.org/10.1103/PhysRevLett.93.180502}{Phys. Rev. Lett. \textbf{93}, 180502 (2004).}

\bibitem{Salart08}
 D. Salart, A. Baas, C. Branciard, N. Gisin, and H. Zbinden,
 \href{http://www.nature.com/nature/journal/v454/n7206/abs/nature07121.html}{Nature (London) \textbf{454}, 861 (2008).}

\bibitem{Dynes09}
 J. F. Dynes, H. Takesue, Z. L. Yuan, A. W. Sharpe, K. Harada, T. Honjo, H. Kamada, O. Tadanaga, Y. Nishida, M. Asobe, and A. J. Shields,
 \href{http://dx.doi.org/10.1364/OE.17.011440}{Opt. Express \textbf{17}, 11440 (2009).}

\bibitem{Inagaki13}
 T. Inagaki, N. Matsuda, O. Tadanaga, M. Asobe, and H. Takesue,
 \href{http://dx.doi.org/10.1364/OE.21.023241}{Opt. Express \textbf{21}, 23241 (2013).}

\bibitem{Acin07}
 A. Ac\'in, A., N. Brunner, N. Gisin, S. Massar, S. Pironio, and V. Scarani,
 \href{http://dx.doi.org/10.1103/PhysRevLett.98.230501}{Phys. Rev. Lett. \textbf{98}, 230501 (2007).}

\bibitem{Larsson14}
 J.-\AA{}. Larsson,
 \href{http://iopscience.iop.org/1751-8121/47/42/424003}{J. Phys. A: Math. Theor. \textbf{47}, 424003 (2014).}

\bibitem{Weihs98}
 G. Weihs, T. Jennewein, C. Simon, H. Weinfurter, and A. Zeilinger,
 \href{http://dx.doi.org/10.1103/PhysRevLett.81.5039}{Phys. Rev. Lett. \textbf{81}, 5039 (1998).}

\bibitem{Rowe01}
 M. A. Rowe, D. Kielpinski, V. Meyer, C. A. Sackett, W. M. Itano, C. Monroe, and D. J. Wineland,
 \href{http://www.nature.com/nature/journal/v409/n6822/abs/409791a0.html}{Nature (London) \textbf{409}, 791 (2001).}

\bibitem{Monroe08}
 D. N. Matsukevich, P. Maunz, D. L. Moehring, S. Olmschenk, and C. Monroe,
 \href{http://dx.doi.org/10.1103/PhysRevLett.100.150404}{Phys. Rev. Lett. \textbf{100}, 150404 (2008).}

\bibitem{Zeilinger13}
 M. Giustina, A. Mech, S. Ramelow,	B. Wittmann, J. Kofler, J. Beyer,  A. Lita, B. Calkins, T. Gerrits, S. W. Nam,	R. Ursin, and A. Zeilinger,
 \href{http://www.nature.com/nature/journal/v497/n7448/full/nature12012.html}{Nature (London) \textbf{497}, 227 (2013).}

\bibitem{Kwiat13}
 B. G. Christensen, K. T. McCusker, J. B. Altepeter, B. Calkins, T. Gerrits, A. E. Lita, A. Miller, L. K. Shalm, Y. Zhang, S. W. Nam, N. Brunner, C. C. W. Lim, N. Gisin, and P. G. Kwiat,
 \href{http://dx.doi.org/10.1103/PhysRevLett.111.130406}{Phys. Rev. Lett. \textbf{111}, 130406 (2013).}

\bibitem{Cuevas13}
 A. Cuevas, G. Carvacho, G. Saavedra, J. Cari\~ne,	W. A. T. Nogueira, M. Figueroa,	A. Cabello, P. Mataloni, G. Lima, and G. B. Xavier,
 \href{http://www.nature.com/ncomms/2013/131129/ncomms3871/full/ncomms3871.html}{Nat. Commun. \textbf{4}, 2871 (2013).}

  \bibitem{Smith2012}
D. H. Smith, G. Gillett, M. P. de Almeida, C. Branciard, A. Fedrizzi, T. J. Weinhold, A. Lita, B. Calkins, T. Gerrits, H. M. Wiseman, S. W. Nam and A. G. White \href{http://www.nature.com/ncomms/journal/v3/n1/full/ncomms1628.html}{Nat. Commun. \textbf{3}, 625 (2012).}

 \bibitem{Bennet2012}
A. J. Bennet, D. A. Evans, D. J. Saunders, C. Branciard, E. G. Cavalcanti, H. M. Wiseman, and G. J. Pryde, \href{http://journals.aps.org/prx/abstract/10.1103/PhysRevX.2.031003}{Phys. Rev. X \textbf{2}, 031003 (2012).}

 \bibitem{Wittmann2012}
B. Wittmann, S. Ramelow, F. Steinlechner, N. K. Langford, N. Brunner, H. M. Wiseman, R. Ursin and A. Zeilinger, \href{http://iopscience.iop.org/1367-2630/14/5/053030}{New J. Phys. \textbf{14} 053030 (2012).}

 \bibitem{Franson89}
 J. D. Franson,
 \href{http://dx.doi.org/10.1103/PhysRevLett.62.2205}{Phys. Rev. Lett. \textbf{62}, 2205 (1989).}

\bibitem{Jogenfors14}
 J. Jogenfors and J.-\AA{}. Larsson,
 \href{http://iopscience.iop.org/1751-8121/47/42/424032/}{J. Phys. A: Math. Theor. \textbf{47}, 424032 (2014).}

 \bibitem{Franson1}
 J. D. Franson, \href{http://journals.aps.org/pra/abstract/10.1103/PhysRevA.61.012105}{Phys. Rev. A \textbf{61}, 12105 (1999).}

 \bibitem{Franson2}
 J. D. Franson, \href{http://journals.aps.org/pra/abstract/10.1103/PhysRevA.80.032119}{Phys. Rev. A \textbf{80}, 032119 (2009).}

\bibitem{Aerts99}
 S. Aerts, P. Kwiat, J.-\AA{}. Larsson, and M. \.{Z}ukowski,
 \href{http://dx.doi.org/10.1103/PhysRevLett.83.2872}{Phys. Rev. Lett. \textbf{83}, 2872 (1999).}

\bibitem{Larsson02}
 J.-\AA{}. Larsson,
 \href{http://dl.acm.org/citation.cfm?id=2011494&CFID=473640422&CFTOKEN=99295618}{Quantum Inf. Comput. \textbf{2}, 434 (2002).}

\bibitem{Jogenfors14b}
 J. Jogenfors, A. M. Elhassan, J. Ahrens, M. Bourennane, and J.-\AA{}. Larsson,
 \href{http://arxiv.org/abs/1411.7222}{\eprint{arXiv:1411.7222}.}

\bibitem{CRVMM09}
 A. Cabello, A. Rossi, G. Vallone, F. De Martini, and P. Mataloni,
 \href{http://dx.doi.org/10.1103/PhysRevLett.102.040401}{Phys. Rev. Lett. \textbf{102}, 040401 (2009).}

\bibitem{LVCCM10}
 G. Lima, G. Vallone, A. Chiuri, A. Cabello, and P. Mataloni,
 \href{http://dx.doi.org/10.1103/PhysRevA.81.040101}{Phys. Rev. A \textbf{81}, 040401(R) (2010).}

\bibitem{Saleh}
 B. E. A. Saleh and M. C. Teich,
 {\em Fundamentals of Photonics}
 (Wiley, New York, 2007).

\bibitem{Meyer-Scott10}
 E. Meyer-Scott, H. H\"{u}bel, A. Fedrizzi, C. Erven, G. Weihs, and T. Jennewein,
 \href{http://scitation.aip.org/content/aip/journal/apl/97/3/10.1063/1.3460920}{Appl. Phys. Lett. \textbf{97}, 031117 (2010).}

\bibitem{Minar08}
 J. Min\'a\v{r}, H. de Riedmatten, C. Simon, H. Zbinden, and N. Gisin,
 \href{http://dx.doi.org/10.1103/PhysRevA.77.052325}{Phys. Rev. A. \textbf{77}, 052325 (2008).}

\bibitem{Xavier11}
 G. B. Xavier and J. P. von der Weid,
 \href{http://dx.doi.org/10.1364/OL.36.001764}{Opt. Lett. \textbf{36}, 1764 (2011).}


\bibitem{CHSH69}
 J. F. Clauser, M. A. Horne, A. Shimony, and R. A. Holt,
 \href{http://dx.doi.org/10.1103/PhysRevLett.23.880}{Phys. Rev. Lett. \textbf{23}, 880 (1969).}

\bibitem{Clauser78}
 J. F. Clauser and A. Shimony,
 \href{http://iopscience.iop.org/0034-4885/41/12/002/}{Rep. Prog. Phys. \textbf{41}, 1881 (1978).}

\end{thebibliography}
\end{document}